\documentclass{ngc4631_x4} 

\usepackage{natbib}
\usepackage{hyperref}
\usepackage[T1]{fontenc}
\usepackage{ae,aecompl}
\usepackage{rotating}
\usepackage{color}
\usepackage{graphicx}
\usepackage{appendix}
\usepackage{longtable}
\usepackage{txfonts}

\begin{document} 

\titlerunning{Spectral and timing variability of NGC 4631 X--4}

\authorrunning{S. Allak et al.} 
\title{Spectral and timing variability of the transient ultraluminous X-ray source NGC 4631 X--4}

 \author{Sinan Allak\inst{1}, Wei Yu\inst{1}, Lorenzo Ducci\inst{1}, Andrea Santangelo\inst{1}, Aysun Akyuz\inst{2}, Faruk Soydugan\inst{3} and Valery Suleimanov\inst{1}}

\institute{Institut für Astronomie und Astrophysik, Sand 1, 72076 
Tübingen, Germany\\
\email{sinan.allak@uni-tuebingen.de}
\and
Department of Physics, University of Çukurova, 01330, Adana, Türkiye
\and
Department of Physics, University of Çanakkale Onsekiz Mart, 17100, Çanakkale, Türkiye
}

 \date{Received ---, ----; accepted ----, ----}

\abstract{
\textit{Context.} Ultraluminous X-ray sources (ULXs) are accreting compact objects exhibiting luminosities above the Eddington limit of stellar-mass black holes, and are key laboratories for studying super-Eddington accretion physics.

\textit{Aims.} We investigate the spectral and timing properties of the transient ULX NGC~4631~X--4 to constrain the nature of the accretion flow and the compact object.

\textit{Methods.} We analyzed archival \textit{Chandra}, \textit{XMM-Newton}, and \textit{Swift}/XRT observations, focusing on long- and short-term variability, spectral evolution, and timing behavior. Spectral fitting was performed using standard absorbed models, while timing analysis was performed to search for variability and periodic or quasi-periodic signals using standard statistical methods.

\textit{Results.} X--4 exhibits strong spectral and flux variability on both short and long timescales, with luminosity changes exceeding two orders of magnitude. The X-ray spectra are adequately described by absorbed multicolor disk blackbody (\textit{diskbb}) and power-law models, with characteristic inner disk temperatures of $kT_{\rm in}\sim0.9$--$1.4$ keV and relatively steep power-law photon indices of $\Gamma \sim 2$--2.4. The source does not follow the standard $L \propto T^{4}$ relation expected for a thin accretion disk. No coherent pulsations or quasi-periodic oscillations are detected. Short-term variability is dominated by aperiodic fluctuations and kilosecond peak-like structures, possibly associated with clumpy winds and geometric effects in a super-Eddington accretion flow.

\textit{Conclusions.} The spectral and timing properties of X--4 are consistent with super-Eddington accretion onto a stellar-mass compact object. The observed variability and spectral evolution can be explained by a combination of a variable accretion rate and a radiatively driven, optically thick wind. The nature of the compact object cannot be uniquely determined from the present data; both neutron star and stellar-mass black hole accretors remain consistent with the observed properties.}


 \maketitle
%

\section{Introduction}

ULXs are off-nuclear point sources with apparent X-ray luminosities exceeding the Eddington limit of a typical stellar-mass black hole ($\sim$10~M$_\odot$), i.e., $L_{\mathrm{X}} \gtrsim 10^{39}\,\mathrm{erg\ s^{-1}}$ \citep{2017ARA&A..55..303K,2021AstBu..76....6F,2023NewAR..9601672K,2023arXiv230200006P}. Their extreme luminosities initially led to the interpretation that ULXs might host intermediate-mass black holes \citep{1999ApJ...519...89C}. However, subsequent observational and theoretical studies have shown that a large fraction of ULXs can instead be explained by super-Eddington accretion onto stellar-mass compact objects \citep{2009MNRAS.397.1836G,2015MNRAS.454.3134M}. As a result, ULXs are now widely recognized as key laboratories for studying the physics of super-Eddington accretion, accretion geometry, and energetic feedback under extreme conditions \citep{2020A&A...643A.171H}.

The discovery of coherent X-ray pulsations in several ULXs has provided direct evidence that neutron stars are capable of sustaining accretion at luminosities exceeding the Eddington limit by up to two orders of magnitude (e.g., \citealt{2014Natur.514..202B,2014Natur.514..198M,2017Sci...355..817I,2018MNRAS.476L..45C,2025arXiv251104282D}). These pulsating ULXs (PULXs) have firmly established the role of strongly magnetized neutron stars within the ULX population and shown that extreme accretion regimes are not limited to black hole systems \citep{2016MNRAS.458L..10K,2016MNRAS.457.1101T}. Nevertheless, X-ray pulsations are often transient or weak and may remain undetected in many systems, leaving the nature of the compact object in the majority of ULXs uncertain. In this context, X-ray variability represents one of the most powerful observational tools for probing the physical properties of ULXs. On long timescales, many ULXs, particularly transient systems, exhibit luminosity variations spanning several orders of magnitude, reflecting substantial changes in the mass accretion rate and accretion flow configuration \citep{2009MNRAS.398.1450K,2017ARA&A..55..303K}. 

Transient ULXs are characterized by episodic outbursts separated by extended low-luminosity or quiescent intervals, providing a unique opportunity to investigate accretion physics across a wide dynamic range in luminosity and spectral appearance \citep{2019ApJ...881...38E}. On shorter timescales, ULXs display a wide range of variability behaviors, from relatively stable emission to pronounced aperiodic fluctuations, which are often strongly energy dependent \citep{2013MNRAS.435.1758S,2015MNRAS.447.3243M}. Within the framework of super-Eddington accretion, such variability is commonly interpreted as arising from a combination of intrinsic fluctuations in the accretion flow and changes in accretion geometry and viewing angle, rather than classical state transitions observed in sub-Eddington Galactic X-ray binaries (XRBs) \citep{2006ARA&A..44...49R}. In particular, optically thick, radiatively driven winds can obscure or reprocess the inner accretion flow, producing complex and non-monotonic hardness--luminosity relations and irregular transitions between harder and softer spectral regimes \citep{2009MNRAS.397.1836G,2015MNRAS.454.3134M}. In transient ULXs, these effects may be further enhanced by variations in the mass accretion rate, wind strength, and line-of-sight obscuration during the outburst evolution \citep{2020A&A...643A.171H}.

\begin{figure}
\resizebox{\hsize}{!}{\includegraphics{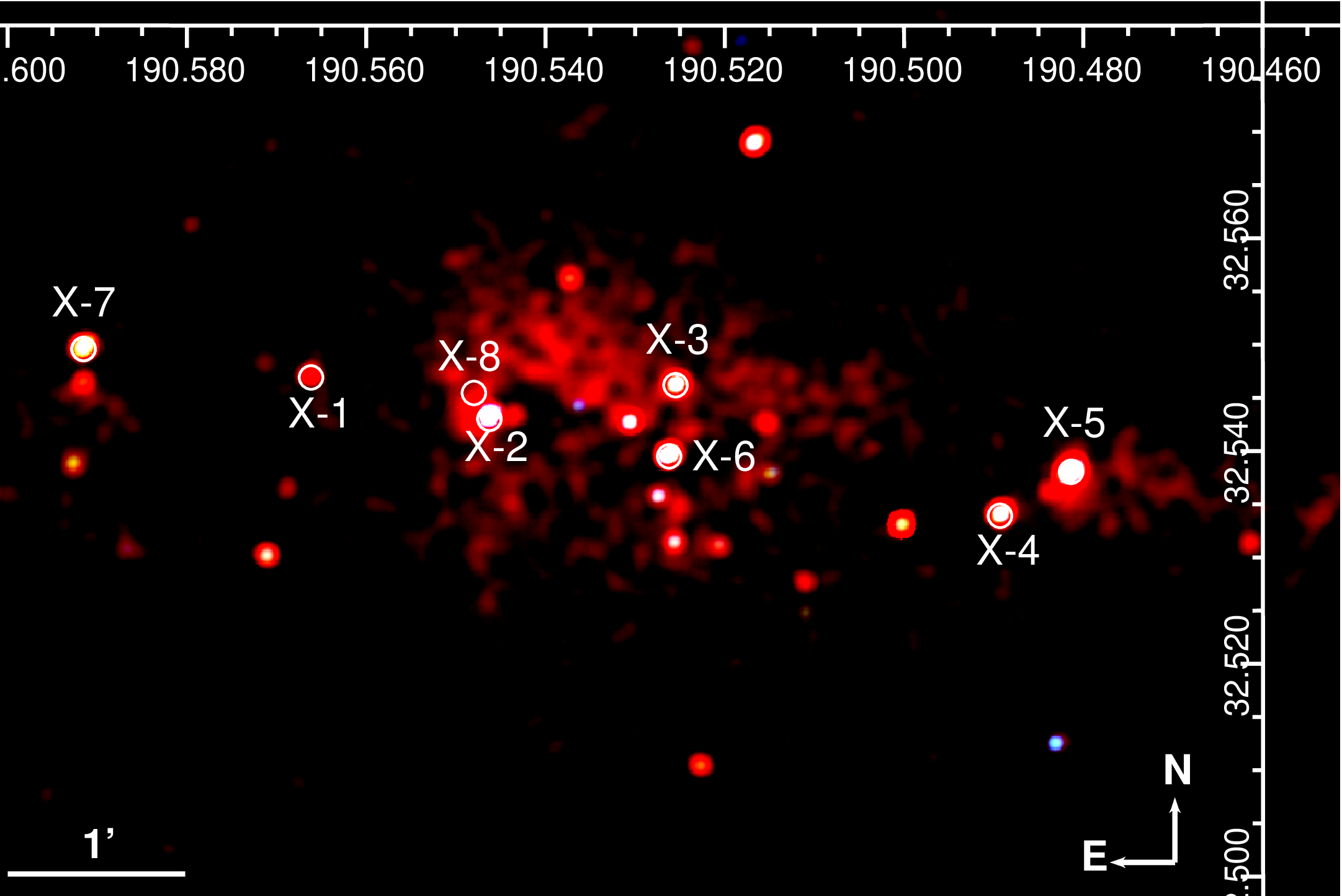}}
\caption{\textit{Chandra} stacked three-color X-ray image of NGC~4631. Red, green, and blue correspond to the 0.5--1.0~keV, 1.0--2.0~keV, and 2.0--8.0~keV energy bands, respectively. The image was smoothed with a Gaussian kernel of $5^{\prime\prime}$. White circles indicate the positions of the ULXs. Source X--1 is classified as a supersoft ULX, X--6 and X--7 as transient ULXs, and X--8 as a pulsating transient ULX. North is up and east is to the left.}
\label{F:rgb}
\end{figure}

Transient X--4 is one of the ULXs in the nearby, edge-on, star-forming spiral galaxy NGC~4631 (the Whale Galaxy), located at a distance of 7.5~Mpc ($1^{\prime\prime} \approx 35$~pc; \citealt{2009ApJ...696..287S}). NGC~4631 is known for its active star formation and prominent extra-planar emission, providing a favorable environment for the formation and evolution of luminous XRBs. The galaxy hosts a rich ULX population (see Fig. \ref{F:rgb}), including X--2, X--3, X--4, and X--5 identified in early \textit{Chandra} and \textit{XMM-Newton} observations \citep{2009ApJ...696..287S}. In addition, the most luminous source, X--1, was reported as a highly variable supersoft ULX with a bolometric luminosity of a few $\times10^{39}$~erg~s$^{-1}$ \citep{2009ApJ...696..287S}. Optical studies of ULXs in NGC~4631 have revealed complex source environments and signatures of strong feedback, including a highly asymmetric bubble nebula around X--4 that is likely powered by jet- or outflow-driven shocks \citep{2023ApJ...946...72G}. More recently, NGC~4631~X--8 has been identified as a pulsating transient ULX \citep{2025ApJ...994L..38D}, while X--6 and X--7 have been classified as transient systems consistent with high-mass XRBs hosting stellar-mass compact objects \citep{2026arXiv260108047A}. These properties make NGC~4631 an excellent laboratory for studying ULX variability and accretion physics.

In this work, we present the first dedicated spectral and timing study of the transient ULX X--4 based on a comprehensive set of archival \textit{Chandra}, \textit{XMM-Newton}, and \textit{Swift}/XRT observations. By combining multi-epoch data, we probe the source variability over a wide range of timescales, from years down to kilosecond intervals. This approach enables us to characterize the spectral evolution, short-term aperiodic variability, and accretion properties of X--4, and to constrain the nature of the accretion flow and the compact object. This paper is organized as follows. In Section \ref{DRA}, we describe the \textit{Chandra}, \textit{XMM-Newton}, and \textit{Swift}/XRT observations of NGC~4631 and outline the data reduction procedures and analysis methods. In Section \ref{RD}, we present and discuss the spectral and timing results, including the short- and long-term variability and spectral evolution, and interpret the results in the context of super-Eddington accretion models. Finally, Section \ref{C} summarizes the main results and conclusions.

\section{Observations, data reduction and analysis} \label{DRA}

NGC~4631 has been observed multiple times by \textit{Chandra}/ACIS-S, \textit{XMM-Newton}/EPIC, and \textit{Swift}/XRT over the past two decades; details are given in Table \ref{tab:obslog}.

\begin{table*}[ht]
\caption{Log of X-ray observations of NGC~4631 used in this work.}
\label{tab:obslog}
\centering
\small
\begin{tabular}{l l l l l c}
\hline
Mission & Label & ObsID & Instrument & Observation Date & Exposure (ks) \\
\hline

Chandra & C1 & 797 & ACIS-S & 2000-04-16 & 59.21 \\

XMM-Newton & XMM1 & 0110900201 & EPIC & 2002-06-28 & 54.80 \\

Swift & -- & 00082263001 & XRT & 2013-11-08 & 7.37 \\
Swift & -- & 00082263002 & XRT & 2013-11-10 & 2.35 \\
Swift & -- & 00082263003 & XRT & 2013-11-16 & 2.92 \\
Swift & -- & 00082263004 & XRT & 2013-11-18 & 0.91 \\
Swift & -- & 00082263005 & XRT & 2013-11-20 & 3.52 \\
Swift & -- & 00082263006 & XRT & 2013-11-21 & 1.99 \\
Swift & -- & 00084441001 & XRT & 2014-10-24 & 1.26 \\
Swift & -- & 00084441003 & XRT & 2018-03-23 & 0.39 \\
Swift & -- & 00084441004 & XRT & 2018-03-26 & 0.13 \\
Swift & -- & 00084441005 & XRT & 2018-05-20 & 0.31 \\
Swift & -- & 00084441006 & XRT & 2018-05-22 & 0.14 \\
Swift & -- & 00084441007 & XRT & 2018-05-23 & 0.20 \\
Swift & -- & 00084441008 & XRT & 2018-11-21 & 0.56 \\
Swift & -- & 00084441009 & XRT & 2020-03-14 & 0.19 \\
Swift & -- & 00084441010 & XRT & 2020-08-11 & 0.75 \\
Swift & -- & 00084441011 & XRT & 2020-11-10 & 0.91 \\
Swift & -- & 00084441012 & XRT & 2020-11-11 & 0.81 \\
Swift & -- & 00084441013 & XRT & 2020-11-14 & 0.63 \\
Swift & -- & 00084441014 & XRT & 2020-11-16 & 0.51 \\
Swift & -- & 00084441015 & XRT & 2020-11-18 & 0.53 \\
Swift & -- & 00084441016 & XRT & 2020-11-22 & 0.61 \\
Swift & -- & 00084441017 & XRT & 2020-11-26 & 0.45 \\
Swift & -- & 00084441018 & XRT & 2021-02-13 & 1.84 \\

XMM-Newton & XMM2 & 0890710101 & EPIC & 2021-12-28 & 33.00 \\

Chandra & C2 & 25777 & ACIS-S & 2022-01-22 & 29.03 \\
Chandra & C3 & 25220 & ACIS-S & 2022-08-02 & 22.77 \\
Chandra & C4 & 26484 & ACIS-S & 2022-08-02 & 18.82 \\
Chandra & C5 & 26485 & ACIS-S & 2022-08-05 & 20.80 \\
Chandra & C6 & 26486 & ACIS-S & 2022-08-06 & 14.88 \\
Chandra & C7 & 26487 & ACIS-S & 2022-08-07 & 14.88 \\

Chandra & C8 & 25782 & ACIS-S & 2023-01-29 & 30.66 \\
Chandra & C9 & 25780 & ACIS-S & 2023-06-16 & 11.92 \\
Chandra & C10 & 25779 & ACIS-S & 2023-07-04 & 19.82 \\
Chandra & C11 & 25778 & ACIS-S & 2023-07-04 & 19.81 \\
Chandra & C12 & 25781 & ACIS-S & 2023-07-18 & 13.60 \\
XMM-Newton & XMM3 & 0943790101 & EPIC & 2025-07-08 & 104.20 \\
\hline
\end{tabular}
\end{table*}

The \textit{Chandra} ACIS-S observations were reduced using the \textit{Chandra} Interactive Analysis of Observations (CIAO; \citealt{2006SPIE.6270E..1VF}) together with the latest calibration database (CALDB v4.11). Level~2 event files were generated with the standard \textit{chandra\_repro} pipeline. Source detection was performed using \textit{CIAO} \textit{wavdetect} on each individual \textit{Chandra} observation and on the combined (merged) image to ensure the robustness of the detected sources. \textit{wavdetect} was run with wavelet scales of 1, 2, 4, and 8 pixels, and a significance threshold of 10$^{-6}$ in the 0.3--10 keV energy band. The source position of X--4 was determined using \textit{wavdetect}, yielding RA = 12:41:57.48 and Dec = +32:32:02.75 (J2000). The source positions derived from wavdetect are fully consistent, within the astrometric uncertainties, with those reported by \cite{2009ApJ...696..287S}. Spectra and light curves were extracted using \textit{specextract} and \textit{dmextract}, respectively. Source counts were extracted from a circular region of radius $4^{\prime\prime}$ centered on the source, while the background was taken from a concentric annulus with inner and outer radii of $6^{\prime\prime}$ and $12^{\prime\prime}$, respectively. The background region was selected to avoid nearby sources and diffuse emission. For the \textit{Chandra} spectra, which have lower count statistics, we grouped the data to 5--15 counts per bin and adopted the C-statistic, appropriate for Poisson-distributed data.

The \textit{XMM-Newton} EPIC data were reduced with the Science Analysis System (\textit{SAS}) v22.0 and the most recent calibration files. Calibrated event lists were produced using \textit{epproc} for EPIC-pn and \textit{emproc} for EPIC-MOS. Time intervals affected by high particle background were excluded using good-time-interval (GTI) filtering based on high-energy light curves. Valid events were selected by requiring $\mathrm{PATTERN} \leq 4$ and $\mathrm{FLAG} == 0$ for pn, and $\mathrm{PATTERN} \leq 12$ and $\mathrm{FLAG} == 0$ for MOS. Source spectra were extracted from a circular region of radius $15^{\prime\prime}$ centered on X--4, while background events were taken from a nearby source-free region of radius $30^{\prime\prime}$ on the same CCD. Response matrices and ancillary response files were generated using \textit{rmfgen} and \textit{arfgen}. Spectra were grouped with \textit{grppha} to a minimum of 20 counts per bin, allowing the use of $\chi^{2}$ statistics. 

The \textit{Swift}/XRT Photon Counting (PC) mode data were processed with \textit{xrtpipeline} in HEASOFT v6.35.2 using the latest calibration files. Event files were further processed and filtered using \textit{XSELECT}. Source and background events were extracted from circular regions with radii of $20^{\prime\prime}$ and $40^{\prime\prime}$, respectively. We considered three energy bands (0.3--2.0~keV, 2.0--10.0~keV, and 0.3--10~keV). For non-detections, $3\sigma$ upper limits were derived assuming Poisson statistics based on the background counts in the source region.

All spectral fitting was performed using XSPEC \citep{Arnaud1996}. We used the 0.3--10~keV energy range for \textit{Chandra} and \textit{Swift}/XRT, and 0.2--12~keV for \textit{XMM-Newton}. We tested several standard spectral models, including single-component models (\textit{diskbb}, power-law), composite models (\textit{diskbb+power-law}, \textit{diskbb+bbodyrad}), and more physically motivated models (\textit{cutoffpl}, \textit{compTT}), all modified by interstellar absorption (\textit{tbabs}). Due to the limited photon statistics, the additional complexity introduced by the \textit{cutoffpl} and \textit{compTT} models did not result in statistically significant improvements to the fits. In addition, key physical parameters in the \textit{compTT} model (e.g., electron temperature and optical depth) remained poorly constrained within their uncertainties. Therefore, we discuss the absorbed power-law and \textit{diskbb} models as the simplest models that provide an adequate description of the available data.

\begin{figure}
\resizebox{\hsize}{!}{\includegraphics{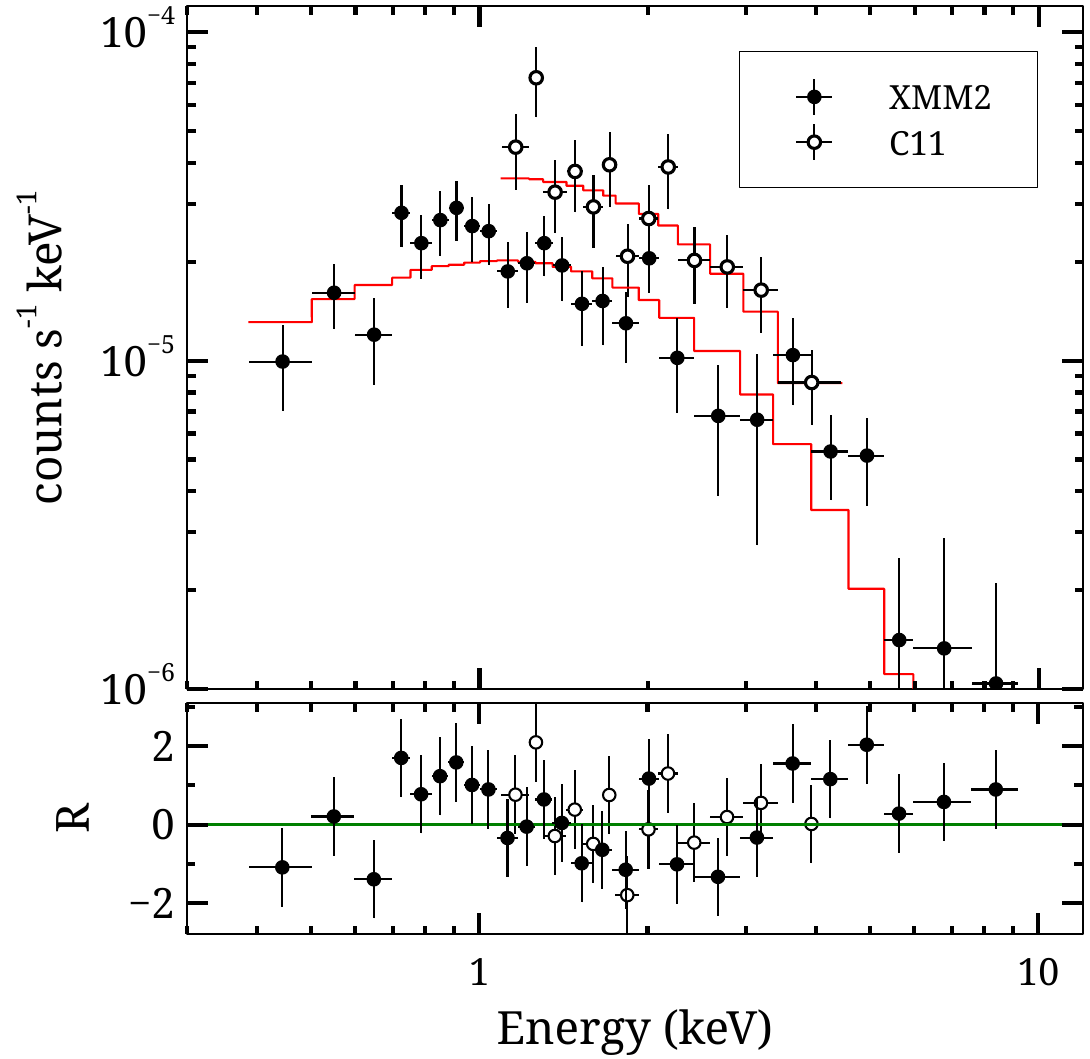}}
\caption{Energy spectra of X--4 obtained with XMM2 (filled circles) and \textit{Chandra} C11 (open circles). The \textit{tbabs*diskbb} is shown as a solid red line. The lower panels show the residuals in units of $R=(\mathrm{data}-\mathrm{model})/\mathrm{error}$, where the error corresponds to the 1$\sigma$ uncertainty of each data point}.
\label{F:spectra}
\end{figure}

\begin{figure}
\resizebox{\hsize}{!}{\includegraphics{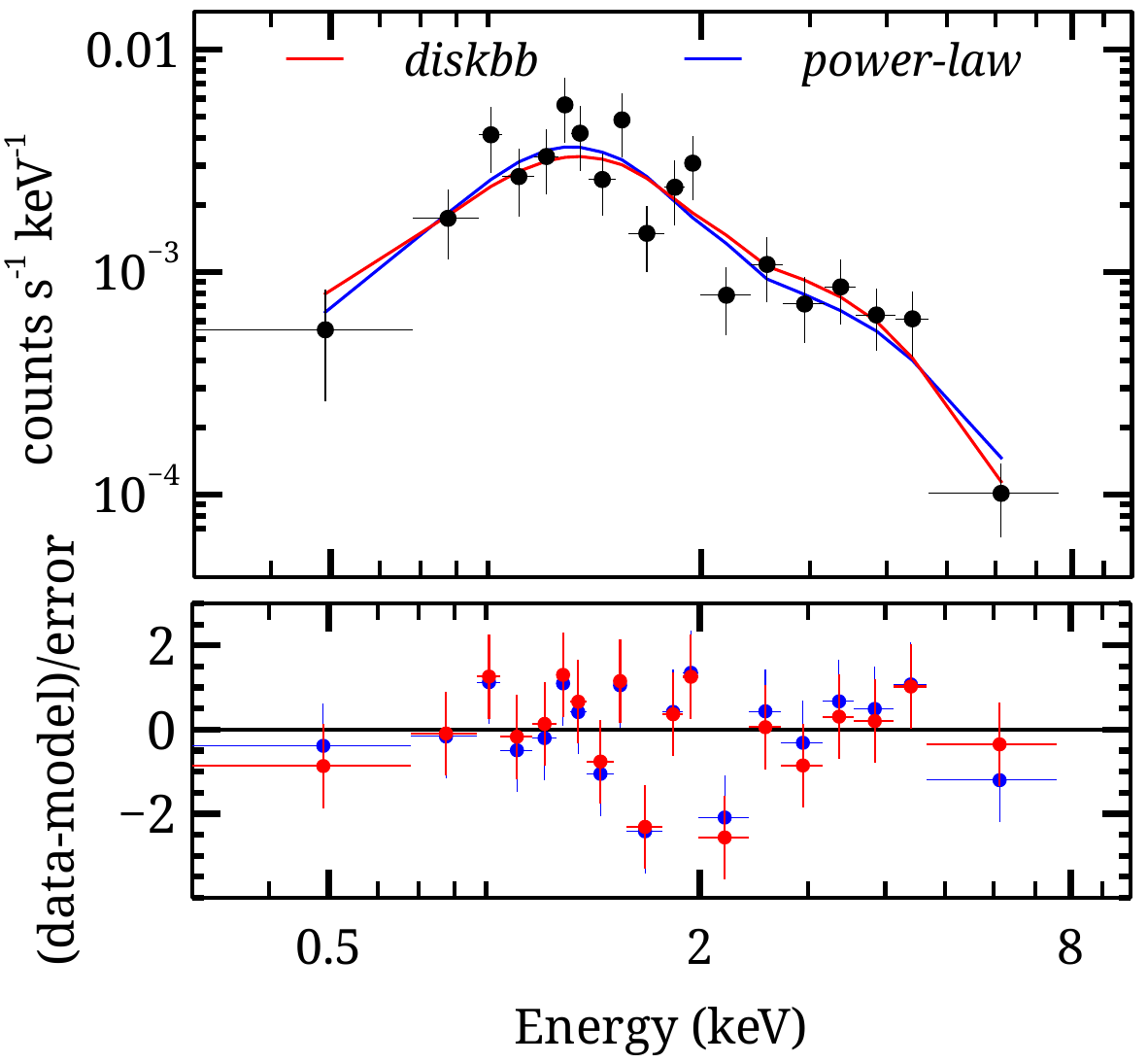}}
\caption{Time-averaged \textit{Swift}/XRT energy spectrum and fit residuals of ULX NGC~4631~X--4 in the 0.3--10 keV energy range. The spectrum is fitted with an absorbed \textit{diskbb} model (red solid line) and, for comparison, an absorbed power-law model (blue solid line).}
\label{F:spectraXRT}
\end{figure}

For the \textit{Chandra} spectra, the absorption column density remained poorly constrained in both the \textit{diskbb} and power-law fits. We therefore fixed $N_{\rm H}$ to the values derived from the best-fit XMM2 spectrum for each respective model. For the \textit{diskbb} fits, $N_{\rm H}$ was fixed to $0.06 \times 10^{22}\ \mathrm{cm^{-2}}$, while for the power-law fits it was fixed to $0.24 \times 10^{22}\ \mathrm{cm^{-2}}$. Only observations with sufficient photon statistics to constrain the spectral parameters reliably were included in the detailed spectral analysis presented in Table~\ref{T:parametr_ch} (C2, C8, C11, C12 and XMM2). The \textit{Chandra} C11 and XMM2 spectra are shown in Fig.~\ref{F:spectra}. Due to the limited photon statistics of the \textit{Swift}/XRT data, the time-averaged spectrum was fitted with simple single-component models, namely an absorbed power-law and an absorbed \textit{diskbb}. Both models provide statistically acceptable fits with physically reasonable parameters. The resulting spectra are shown in Fig.~\ref{F:spectraXRT}, and the best-fit parameters are listed in Table~\ref{T:specraXRT}.

\begin{table*}[ht]
\centering
\caption{Best-fit parameters of the \textit{tbabs*diskbb} and \textit{tbabs*power-law} models for X--4.}
\begin{tabular}{llll cc c c c ccc}
\hline
Parameter & Unit/Observations & C2 & C8 & C11 & C12 & XMM2 \\
\hline
&&&\textit{tbabs*diskbb} \\
$N_{\rm H}$\textsuperscript{(1)} & $10^{22}$ cm$^{-2}$ & 0.06 (fixed) & 0.06 (fixed) & 0.06 (fixed) & 0.06 (fixed) & $0.06^{+0.03}_{-0.02}$ \\
$T_{\rm in}$\textsuperscript{(2)} & keV & $1.26^{+0.41}_{-0.28}$ & $0.93^{+0.51}_{-0.25}$ & $0.90^{+0.14}_{-0.11}$ & $1.05^{+0.18}_{-0.10}$ & $1.32^{+0.18}_{-0.16}$ \\
Norm\textsuperscript{(3)} & $\times10^{-3}$ & $1.88^{+0.88}_{-0.36}$ & $2.05^{+2.46}_{-1.92}$ & $13.96^{+6.12}_{-1.27}$ & $6.35^{+3.76}_{-1.26}$ & $2.24^{+1.13}_{-0.80}$ \\
Observed flux\textsuperscript{(4)} & $10^{-13}$ erg cm$^{-2}$ s$^{-1}$ & $0.72^{+0.16}_{-0.04}$ & $0.27^{+0.04}_{-0.01}$ & $1.68^{+0.11}_{-0.06}$ & $1.37^{+0.12}_{-0.08}$ & $1.28^{+0.24}_{-0.16}$ \\
Luminosity\textsuperscript{(5)} & $10^{39}$ erg s$^{-1}$ & $0.53^{+0.08}_{-0.02}$ & $0.20^{+0.04}_{-0.02}$ & $1.27^{+0.18}_{-0.08}$ & $1.02^{+0.06}_{-0.04}$ & $0.66^{+0.18}_{-0.12}$ \\
Statistics/dof\textsuperscript{(6)} & & 33.62/12 & 10.12/10 & 11.54/11 & 8.91/9 & 31.45/23 \\
Goodness (\%)\textsuperscript{(7)} &&44&29&24&44&73\\

\hline
&&&\textit{tbabs*power-law} \\
$N_{\rm H}$\textsuperscript{(1)} & $10^{22}$ cm$^{-2}$ & 0.24 (fixed) & 0.24 (fixed) & 0.24 (fixed) & 0.24 (fixed) & $0.24^{+0.11}_{-0.08}$ \\

$\Gamma$\textsuperscript{} & & $2.05^{+0.36}_{-0.33}$ & $2.15^{+0.61}_{-0.51}$ & $2.34^{+0.34}_{-0.31}$ & $2.06^{+0.40}_{-0.39}$ & $2.41^{+0.50}_{-0.41}$ \\

Norm\textsuperscript{} & $\times10^{-5}$ & $2.13^{+0.68}_{-0.53}$ & $0.97^{+0.56}_{-0.32}$ & $6.41^{+1.69}_{-1.47}$ & $4.52^{+1.76}_{-1.34}$ & $4.10^{+1.63}_{-1.08}$ \\

Observed flux\textsuperscript{(4)} & $10^{-13}$ erg cm$^{-2}$ s$^{-1}$ & $1.01^{+0.22}_{-0.11}$ & $0.42^{+0.06}_{-0.02}$ & $2.47^{+0.15}_{-0.08}$ & $2.04^{+0.11}_{-0.06}$ & $1.04^{+0.12}_{-0.08}$ \\

Luminosity\textsuperscript{(5)} & $10^{39}$ erg s$^{-1}$ & $0.80^{+0.12}_{-0.02}$ & $0.34^{+0.08}_{-0.04}$ & $2.13^{+0.16}_{-0.06}$ & $1.65^{+0.08}_{-0.02}$ & $1.30^{+0.10}_{-0.04}$ \\

Statistics/dof\textsuperscript{(6)} & & 19.89/12 & 8.94/10 & 10.68/11 & 8.08/9 & 20.42/23 \\

Goodness (\%)\textsuperscript{(7)} &&24&29&52&37&36\\ 

\hline
\end{tabular}
\tablefoot{Spectral parameters are within the 90\% confidence interval.
\textsuperscript{(1)} Equivalent hydrogen column density. For the \textit{Chandra} observations, $N_{\mathrm{H}}$ was fixed to the values derived from the XMM2 fits for each respective model.
\textsuperscript{(2)} Inner disk temperature from the \textit{diskbb} component. 
\textsuperscript{(3)} Normalization of the \textit{diskbb} model: $\mathrm{Norm} = \big(\tfrac{R_{\rm in}/{\rm km}}{D_{10}}\big)^2 \cos\theta$. 
\textsuperscript{(4)} Observed (absorbed) flux in the 0.3--10 keV band. 
\textsuperscript{(5)} Unabsorbed luminosity in the 0.3--10 keV band. All uncertainties are quoted at the 90\% confidence level. 
\textsuperscript{(6)} For C2, C8, and C12, the spectra were grouped to a minimum of 10 counts per bin and fitted using the C-statistic (Cstat/dof). The C11 spectrum was grouped to 15 counts per bin and fitted using the C-statistic, while the XMM2 spectrum was grouped to a minimum of 20 counts per bin and fitted using $\chi^{2}$/dof.}
\textsuperscript{(7)} Goodness values indicate the percentage of simulations with a lower test statistic than the data; values near 50\% correspond to an acceptable fit.
\label{T:parametr_ch}
\end{table*}

\begin{table}
\caption{Best-fit parameters of X--4 obtained from the \textit{tbabs*powerlaw} and \textit{tbabs*diskbb} models using the time-averaged \textit{Swift}/XRT spectrum.}
\label{T:specraXRT}
\begin{tabular}{cccc}
\hline
Parameters & Units & Values & Values \\
 & & (po) & (diskbb) \\
\hline

N$_{\mathrm{H}}$ (a) & $10^{22}$ ($\mathrm{cm^{-2}}$)
& $0.28^{+0.15}_{-0.12}$
& $0.12^{+0.10}_{-0.08}$ \\

${\Gamma}$ &
& $1.86^{+0.46}_{-0.24}$
& \\

$N_{\rm norm}$
& $10^{-5}$ (keV$^{-1}$ cm$^{-2}$ s$^{-1}$)
& $8.32^{+0.64}_{-0.46}$
& \\

$T_{\rm in}$ & keV
&
& $1.52^{+0.41}_{-0.28}$ \\

$N_{\rm norm}$
& $10^{-3}\big(\tfrac{R_{\rm in}/{\rm km}}{D_{10}}\big)^2 \cos\theta$
&
& $4.53^{+0.91}_{-0.24}$ \\

F$_{\mathrm{X}}$(b)
& $10^{-13}$ erg cm$^{-2}$ s$^{-1}$
& $3.95^{+0.09}_{-0.07}$
& $3.48^{+0.08}_{-0.06}$ \\

L$_{\mathrm{X}}$(c)
& $10^{39}$ erg s$^{-1}$
& $4.10^{+0.42}_{-0.21}$
& $2.83^{+0.64}_{-0.28}$ \\

$\chi^{2}$/dof
&
& 19.80/16
& 20.60/16 \\

\hline
\end{tabular}

\tablefoot{
(a) Intrinsic hydrogen column density ($N_{\mathrm{H}}$) obtained from the spectral fit.
(b) Absorbed flux in the 0.3--10 keV band.
(c) Unabsorbed luminosity in the 0.3--10 keV band, assuming a distance of 7.5 Mpc.
All uncertainties are quoted at the 90\% confidence level.}
\end{table}

\subsection{Short-term variability}

To examine the short-term variability, we extracted background-subtracted light curves from all available \textit{Chandra} and \textit{XMM-Newton} observations in different energy ranges (e.g., 0.3--10 keV, 0.3--2 keV, 2--10 keV). For \textit{Chandra}, the CIAO tool \textit{dmextract} was employed, while the \textit{SAS} task \textit{evselect} was used for \textit{XMM-Newton}. In both cases, barycentric corrections were applied to the event lists using the \textit{axbary} (for \textit{Chandra}) and \textit{barycen} (for \textit{XMM-Newton}) utilities to transform photon arrival times to the solar system barycenter, thereby removing delays due to the orbital motion of the Earth and the respective observatories. The timing analysis was performed to search for periodic, quasi-periodic (QPO), and pulsar-like signals using standard \textit{HEASOFT} tools. Power Density Spectra (PDS) were computed using the \textit{XRONOS} package. In addition, the Lomb--Scargle periodogram was applied to search for periodicities in unevenly sampled data, while the $Z^2_n$ test was used on event data to search for coherent pulsations. No statistically significant periodic variability, including coherent pulsations or QPOs, was detected in any of the observations.

\begin{figure}
\resizebox{\hsize}{!}{\includegraphics{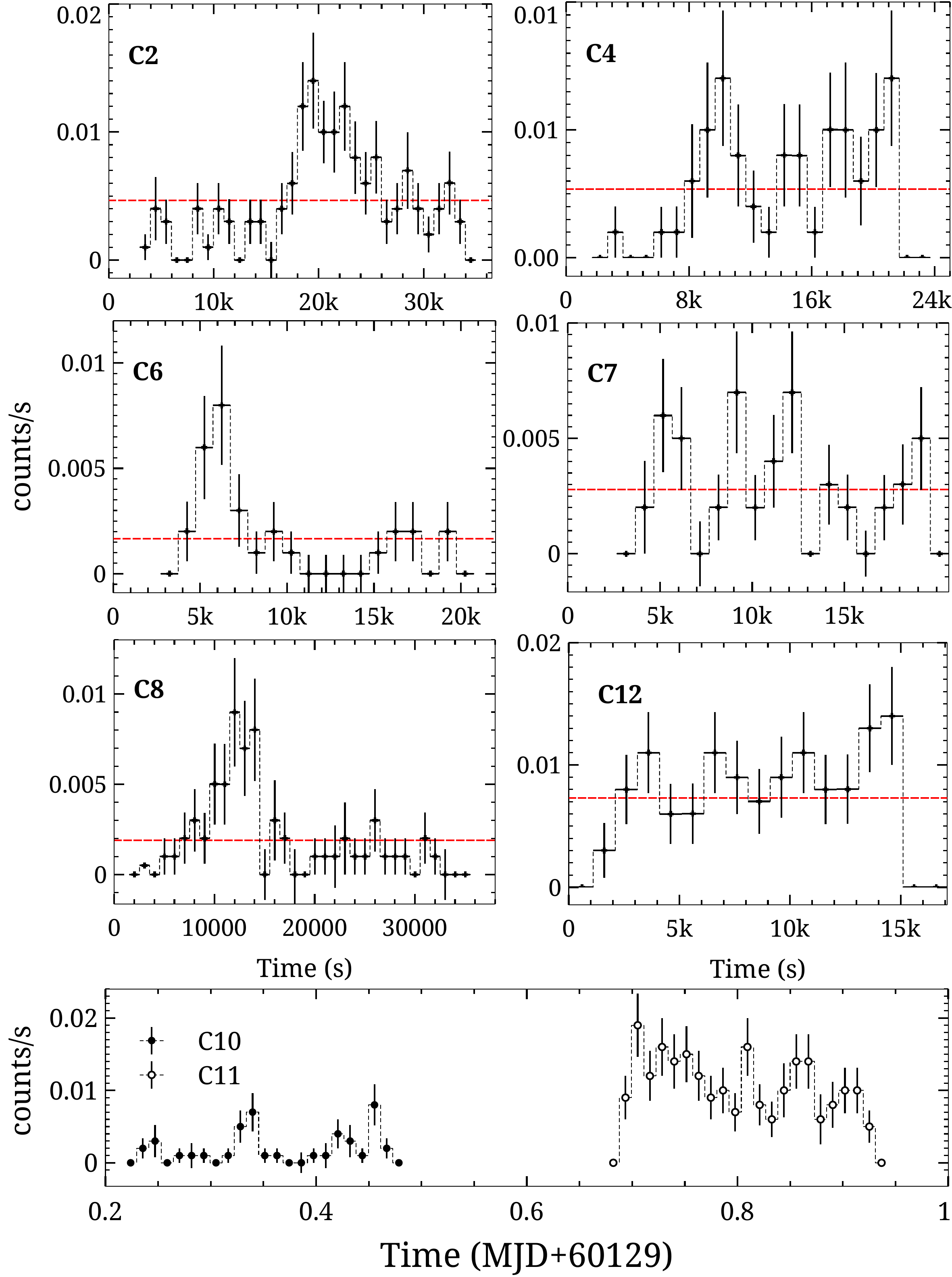}}
\caption{Chandra X-ray light curves of X--4 in NGC\,4631. Each panel shows the background-subtracted count rates in the 0.3--10 keV energy range as a function of time for individual observations (C2, C4, C6, C7, C8, C10, C11, and C12). The red dashed lines indicate the average count rate in each observation. A time bin size of 1000 s was used for all light curves. Error bars represent 1$\sigma$ uncertainties.}
\label{lc_chandra}
\end{figure}

\begin{figure}
\resizebox{\hsize}{!}{\includegraphics{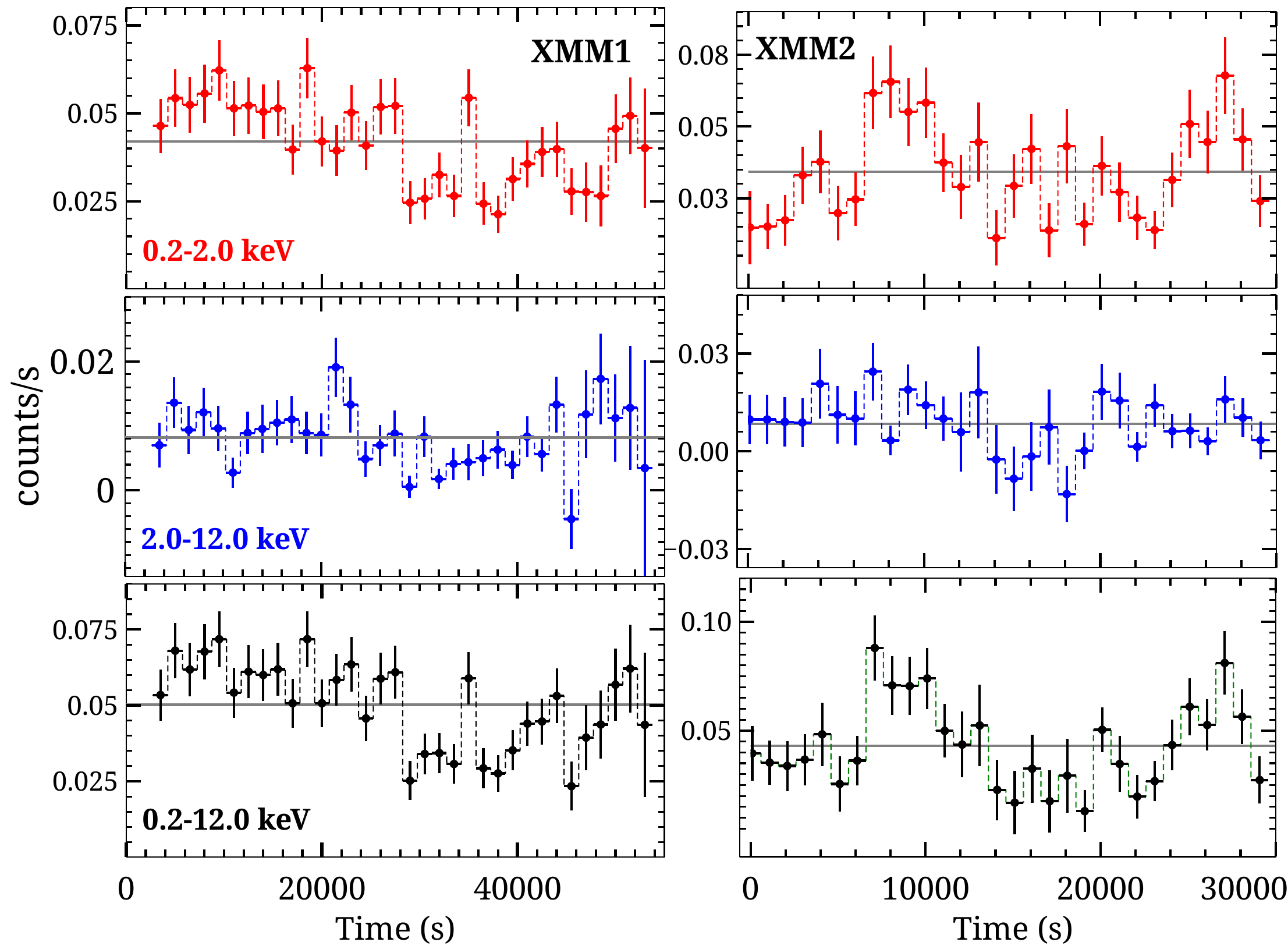}}
\caption{Background-subtracted \textit{XMM-Newton} EPIC-pn light curves of the ULX NGC 4631 X--4 for the two observations analyzed in this work (left: XMM1; right: XMM2). From top to bottom we show the soft (0.2--2.0~keV; red), hard (2.0--12.0~keV; blue), and total (0.2--12.0~keV; black) energy bands. Error bars denote $1\sigma$ uncertainties. The horizontal gray solid lines indicate the mean count rate in each panel.}
\label{lc_xmm}
\end{figure}

Moreover, the light curves were generated using bin sizes of 50, 100, 250, 500, and 1000 s to probe variability over a broad range of timescales and to assess the robustness of the timing analysis against binning effects, given the limited photon statistics. Consistent with this result, no significant periodic signal was found in the binned light-curve analysis; however, peak-like features were identified in the \textit{Chandra} C2, C6, C8, and C10 observations (see Fig.~\ref{lc_chandra}). The peaks observed in the corresponding light-curve profiles are suggestive of short-timescale variability, although their statistical significance is limited by low count statistics. Due to the limited photon statistics in C3, C5, and C9, the data do not allow for a reliable variability analysis, and therefore these observations are not considered further in the timing discussion. To verify whether these features could be instrumental or related to data reduction, we examined all available observations and nearby sources, including the closest ULX X-5. No similar variability patterns were detected in ULX X-5 or other sources, indicating that the observed features are intrinsic to X--4. In addition, as shown in the background-subtracted \textit{Chandra} X-ray light curves (see Fig.~\ref{lc_chandra}), X--4 exhibits markedly different behavior in the two observations obtained on the same day. While no distinct peak-like profiles are present in the \textit{XMM-Newton} observations, the light curves nonetheless reveal clear variability in the X-ray emission (see Fig.~\ref{lc_xmm}).

To investigate the presence of intrinsic short-term variability, we tested the background-subtracted X-ray light curves against a constant-emission model. We fitted the light curves with a constant model and evaluated the goodness of fit using a $\chi^2$ test. Variability is considered statistically significant for $p < 0.01$, values in the range $0.01 \le p < 0.05$ are classified as likely variable, and $p \ge 0.05$ indicates consistency with a constant emission level. The results are summarized in Table~\ref{T:var_test}. Because the $\chi^{2}$ test assumes Gaussian statistics and is sensitive to binning, we additionally applied the \textit{glvary} algorithm implemented in CIAO, which is based on the Gregory--Loredo Bayesian method and operates directly on unbinned photon arrival times. This method compares the likelihood of a constant model with that of piecewise-constant variability models. The algorithm returns both the variability probability (PROB) and the variability index (VI), ranging from 0 (consistent with constant emission) to 10 (highly variable). The \textit{glvary} analysis was applied to all \textit{Chandra} observations, and the corresponding $\chi^{2}$ results and variability indices are listed in Table~\ref{T:var_test}.

\begin{table}
\caption{Results of the $\chi^{2}$ variability test applied to the \textit{Chandra} and \textit{XMM-Newton} observations of ULX X--4.}
\label{T:var_test}
\centering
\begin{tabular}{lcccc}
\hline
Data label & $p(\chi^2)$ & $\chi^2$ cls & VI & GLV cls \\
\hline
\multicolumn{5}{c}{\textit{Chandra}} \\
\hline
C1 & 0.95 & Constant &0& Not variable
\\
C2 & $10^{-4}$ & Var & 8& Variable\\
C3 & 0.99 & Constant & 2& Not variable \\
C4 & 0.15 & Constant & 7& Variable\\
C5 & 0.45 & Constant & 0& Not variable \\
C6 & 0.16 & Constant & 7& Variable
\\
C7 & 0.11 & Constant &1& Not variable
\\
C8 & 0.03 & Likely var &8& Variable\\
C9 & -- & -- && Unconstrained \\
C10 & 0.51 & Constant &6& Variable \\
C11 & 0.13 & Constant &0& Not variable

 \\
C12 & 0.41 & Constant &0& Not variable
 \\
\hline
\multicolumn{5}{c}{\textit{XMM-Newton}} \\
\hline
XMM1 & $10^{-9}$ & Var & -- & --\\
XMM2 & $10^{-4}$ & Var & -- &--\\
XMM3 & -- & -- & -- & Unconstrained \\
\hline
\end{tabular}
\tablefoot{The column $p(\chi^2)$ gives the probability of constancy from the $\chi^2$ test. VI denotes the variability index derived from the \textit{glvary} algorithm. Variability labels are defined as follows: Var. ($p < 0.01$), Likely ($0.01 \le p < 0.05$), Constant ($p \ge 0.05$). GLV cls indicates the classification from the \textit{glvary} test. Unconstrained denotes cases where variability could not be constrained due to limited photon statistics.}
\end{table}

To further test for variability in the low-count regime, we applied a Bayesian Blocks change-point analysis to the unbinned photon arrival times using the \textit{astropy} implementation \citep{2013ApJ...764..167S}. Photon arrival times were extracted from both the source and background regions, and the false-alarm probability parameter was explored in the range $p_{0}=0.01$--$0.1$. Variability was considered significant when multiple blocks were detected in the source light curve while the background remained consistent with a single block. In cases where the $\chi^{2}$ test was consistent with a constant count rate but \textit{glvary} suggested variability, the Bayesian Blocks analysis provided an independent consistency check on kilosecond timescales. The Bayesian Blocks analysis identified multiple statistically significant blocks in observations C2, C6, and C8, supporting the presence of short-timescale variability and the peak-like structures observed in the corresponding light curves. In the remaining observations, the photon statistics were insufficient to identify significant change points, and the light curves were broadly consistent with a single-block (constant) description.

\subsection{Long-term X-ray variability}

To investigate the long-term X-ray variability of X--4, we derived absorbed fluxes through spectral modeling in XSPEC for the \textit{Chandra} and \textit{XMM-Newton} observations with sufficient photon statistics. For the \textit{Swift}/XRT observations, count rates were converted to absorbed fluxes using the web-based PIMMS tool. To ensure consistency across the different instruments, a common reference spectral model was adopted for all count-rate-to-flux conversions and hardness-ratio calculations. We used the best-quality spectrum obtained from the XMM2 observation and adopted the best-fitting absorbed power-law model, with $N_{\rm H}=0.2\times10^{22}\ \mathrm{cm^{-2}}$ and $\Gamma=2.4$. For the low-count \textit{Chandra} observations where spectral fitting was not feasible, the absorbed fluxes were estimated using the CIAO \textit{srcflux} task assuming the same reference spectral model.

\begin{figure*}
\begin{center}
\includegraphics[angle=0,scale=0.45]{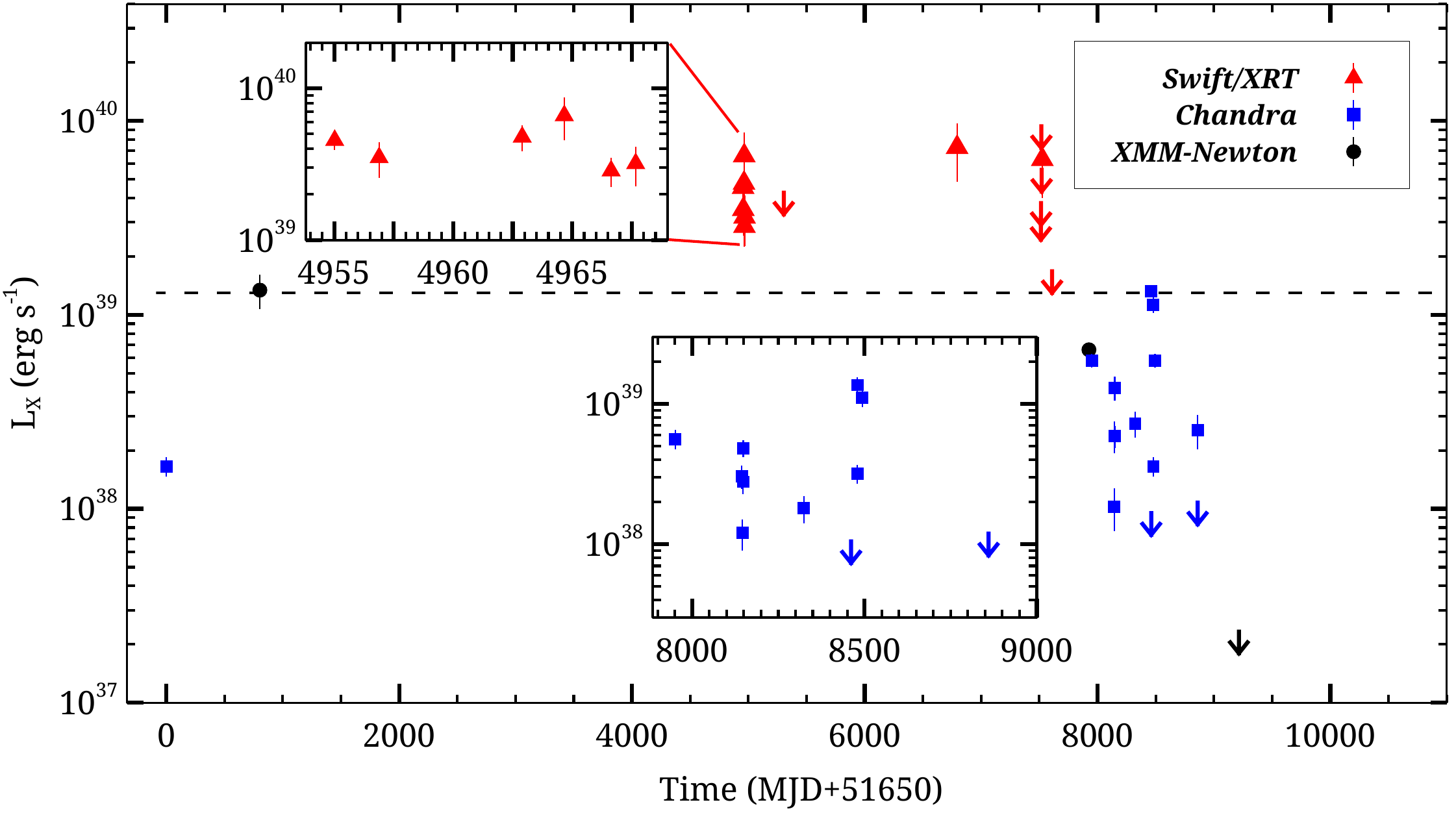}
\caption{Long-term X-ray luminosity evolution of NGC\,4631 X--4 based on \textit{Swift}/XRT (red triangles), \textit{Chandra} (blue squares), and \textit{XMM-Newton} (black circles) observations in the 0.3--10 keV energy band. Luminosities were derived from unabsorbed fluxes assuming a distance of 7.5\,Mpc. The dashed horizontal line indicates the Eddington luminosity for a $10\,M_{\odot}$ black hole, $L_{\rm Edd}=1.3\times10^{39}\,\mathrm{erg\,s^{-1}}$. Downward arrows indicate $3\sigma$ upper limits. Insets show zoomed views of the closely spaced \textit{Swift}/XRT and \textit{Chandra} observations to improve readability. Time is given relative to MJD\,51650, corresponding to 2000 Apr 16.}
\label{F:long_term}
\end{center}
\end{figure*}

\begin{figure*}
\begin{center}
\includegraphics[angle=0,scale=0.45]{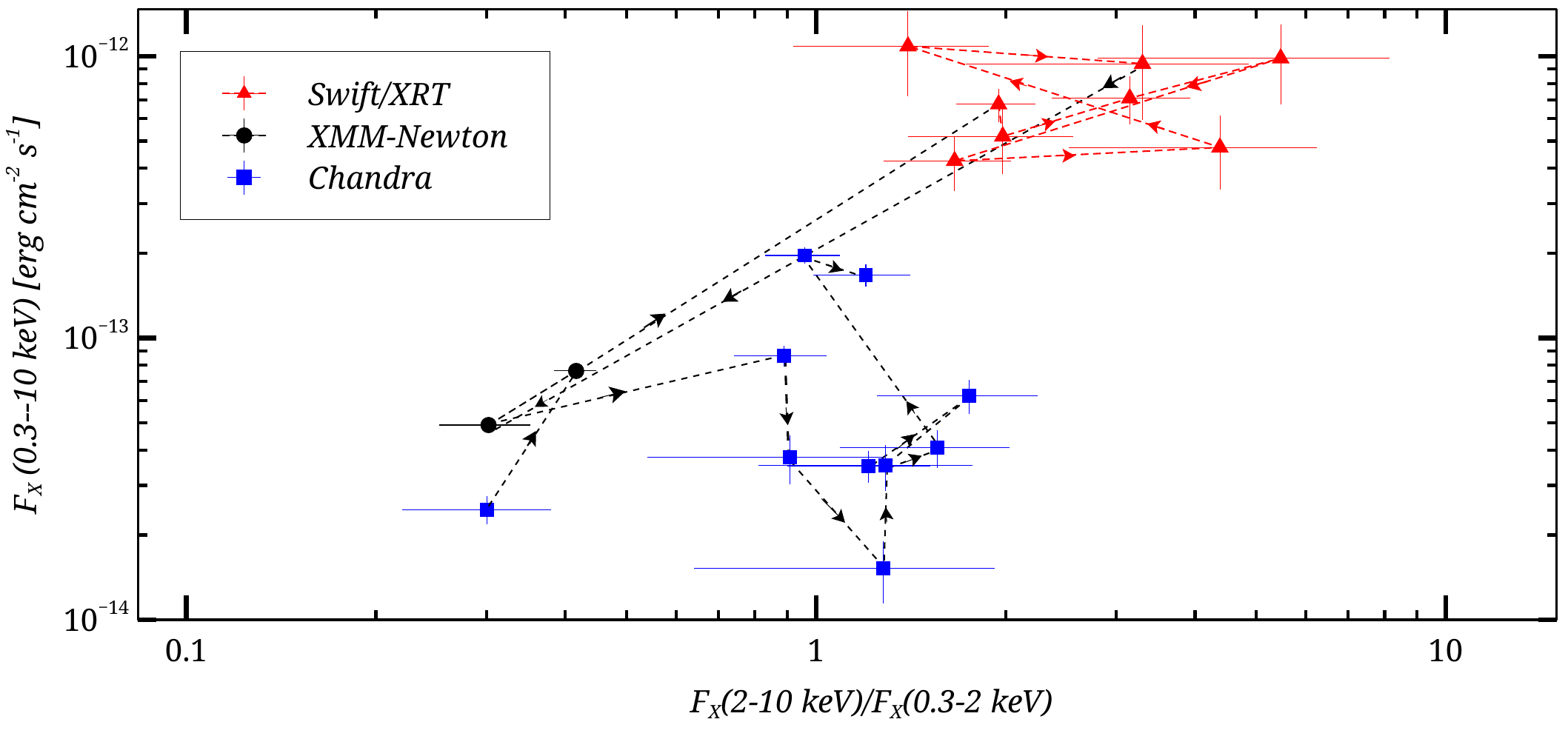}
\caption{Hardness--intensity diagrams of NGC\,4631 X--4 derived from the \textit{Swift}/XRT (red triangles), \textit{XMM-Newton} (black circles), and \textit{Chandra} (blue squares) observations. The hardness ratio is defined as the absorbed flux ratio $F_{X}(2$--$10\,{\rm keV})/F_{X}(0.3$--$2\,{\rm keV})$, while the vertical axis shows the absorbed 0.3--10 keV flux. Dashed arrows connect consecutive observations in chronological order, illustrating the temporal evolution of the source. For clarity, the arrows corresponding to the \textit{Swift}/XRT observations are shown in red, while those for the \textit{Chandra} and \textit{XMM-Newton} observations are shown in black. Error bars represent 1$\sigma$ uncertainties. Both axes are shown on logarithmic scales.}
\label{F:hardness}
\end{center}
\end{figure*}

For observations in which X--4 was not detected, $3\sigma$ upper limits were derived based on Poisson statistics within the source extraction region, accounting for the local background contribution. In the third \textit{XMM-Newton} observation (XMM3), X--4 is not significantly detected in the combined EPIC data, and we derived a $3\sigma$ upper limit of $4.63\times10^{-4}$ counts~s$^{-1}$ (pn), which was converted to flux assuming the same reference spectral model adopted for the flux conversions and hardness-ratio calculations. All unabsorbed fluxes and upper limits were converted to luminosities assuming a distance of 7.5~Mpc in the 0.3--10~keV band. To investigate the spectral evolution of the transient X--4, we constructed hardness ratio (HR) versus time and hardness--intensity diagrams by combining all available observations. The hardness ratio (HR) is defined as the absorbed flux ratio between the hard (2--10~keV) and soft (0.3--2~keV) energy bands, $F_{X}(2$--$10\,{\rm keV})/F_{X}(0.3$--$2\,{\rm keV})$. The long-term X-ray light curve, hardness evolution, and hardness--intensity diagrams are shown in Figs.~\ref{F:long_term} and \ref{F:hardness}, respectively.

\section{Results and discussion} \label{RD}

\subsection{Spectral evolution} \label{Section 3.1}

The equivalent hydrogen column densities derived from the spectral fits are generally modest, although values above the Galactic absorption are obtained in several observations and models. For the \textit{Chandra} spectra, $N_{\rm H}$ remained unconstrained and was therefore fixed to the values derived from the XMM2 spectral fits for the respective models. The XMM2 observation yields moderate absorption column densities of $N_{\rm H}=0.06^{+0.03}_{-0.02}\times10^{22}$ cm$^{-2}$ for the \textit{diskbb} model and $N_{\rm H}=0.24^{+0.11}_{-0.08}\times10^{22}$ cm$^{-2}$ for the power-law model. The time-averaged \textit{Swift}/XRT spectrum yields somewhat higher absorption column densities, with $N_{\rm H}\approx0.12\times10^{22}$ cm$^{-2}$ for the \textit{diskbb} model and $N_{\rm H}\approx0.28\times10^{22}$ cm$^{-2}$ for the power-law model. Within the uncertainties, these $N_{\rm H}$ values are consistent with those measured from the XMM2 observation and therefore do not provide evidence for statistically significant variability in the intrinsic absorption.

The \textit{Chandra} and \textit{XMM-Newton} energy spectra of X--4 can be adequately described by both absorbed \textit{diskbb} and power-law models. The corresponding best-fit spectral parameters are listed in Table~\ref{T:parametr_ch} for a subset of observations (C2, C8, C11, C12, and XMM2) selected based on sufficient photon statistics. The remaining observations do not provide sufficient counts to constrain the spectral parameters reliably and are therefore not included in the Table~\ref{T:parametr_ch}. Although the absorbed \textit{diskbb} and power-law models provide statistically acceptable fits to the available spectra, the current photon statistics do not allow us to robustly constrain the preferred spectral model. As shown in Table \ref{T:parametr_ch}, some observations are better represented by the \textit{diskbb} model, while others show slightly improved fit statistics with the power-law model. Nevertheless, the power-law fits generally yield relatively steep photon indices ($\Gamma \sim 2$--2.4), suggesting that the spectra are not dominated by a hard non-thermal continuum. Within the framework of super-Eddington accretion, such spectra may be associated with thermal emission and reprocessing in an optically thick inner flow and radiatively driven wind, rather than arising purely from non-thermal emission \citep{2009MNRAS.397.1836G}. For this reason, we adopt the \textit{diskbb} model as a phenomenological reference model to characterize the spectral evolution and investigate the luminosity--temperature relation of X--4. We emphasize, however, that the \textit{diskbb} model should be regarded primarily as a phenomenological description of the observed spectral shape in the super-Eddington regime.

The broadband spectral fits of X--4 with the \textit{diskbb} model yield relatively high characteristic temperatures of $kT_{\rm in} \sim 0.9$--1.3 keV (Table~\ref{T:parametr_ch}). Such temperatures are difficult to reconcile with the low disk temperatures expected for sub-Eddington accretion onto intermediate-mass black holes, which would predict significantly cooler disks ($kT_{\rm in} \lesssim 0.3$ keV; see e.g. \citealt{2003ApJ...585L..37M,2004ApJ...614L.117M}). Instead, the observed temperatures are broadly consistent with those commonly observed in stellar-mass compact objects accreting at high rates \citep{2000ApJ...535..632M}. \cite{2009ApJ...696..287S} reported that X--4 displayed a harder ULX-like spectrum during the XMM1 observation, well described by a \textit{power-law + diskbb} model with an unabsorbed luminosity of $\approx 2 \times 10^{39}$ erg s$^{-1}$. In contrast, the earlier \textit{Chandra} observation showed a much softer spectrum consistent with thermal plasma emission, suggesting possible spectral variability associated with the transient nature of the source. In addition, the time-averaged \textit{Swift}/XRT spectrum is adequately fitted by either a power-law model ($\Gamma = 1.83^{+0.42}_{-0.35}$) or a hot disk component ($T_{\rm in} = 1.45^{+0.38}_{-0.28}$ keV), with statistically comparable fits.

\begin{figure}
\resizebox{\hsize}{!}{\includegraphics{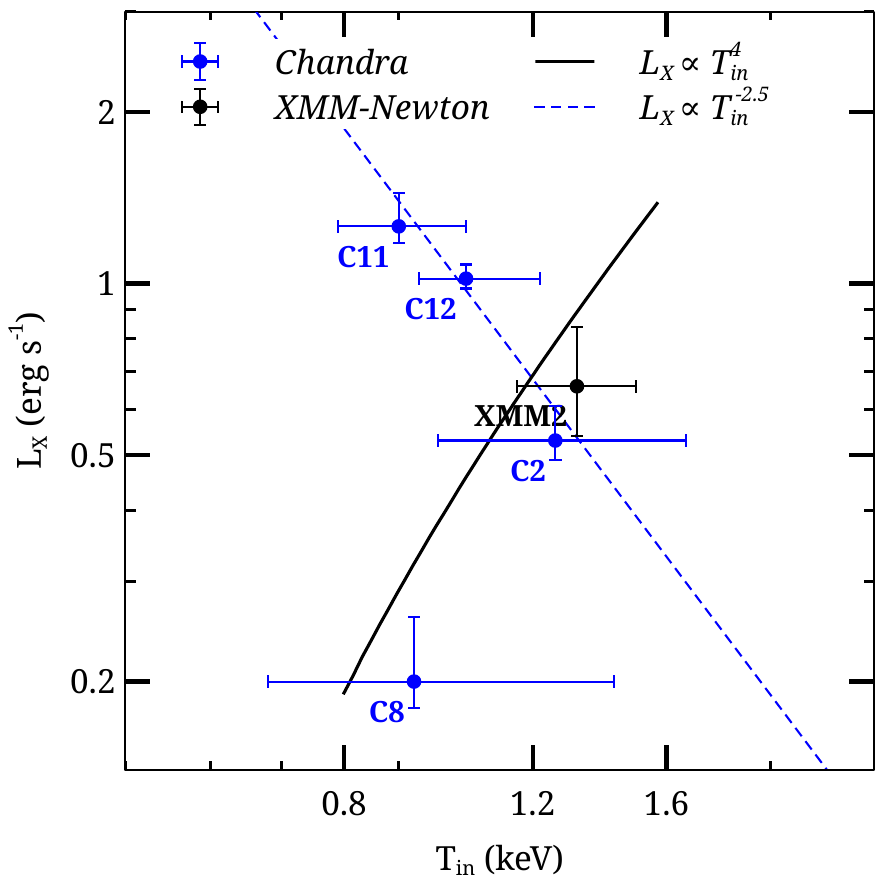}}
\caption{Luminosity--temperature ($L_{\rm X}$--$T_{\rm in}$) relation for X--4 derived from the \textit{tbabs*diskbb} spectral fits. The blue filled circles correspond to the individual \textit{Chandra} observations, while the black filled circle represents the \textit{XMM-Newton} observation. The error bars indicate the 90\% confidence intervals for both parameters. The dashed blue line represents a $L \propto T^{-2.5}$ relation overplotted on the data}. The black solid line represents the theoretical $L_{\rm X} \propto T_{\rm in}^{4}$ relation expected for a standard multicolor disk.
\label{Lx_Tin}
\end{figure}

To further investigate the spectral evolution of X--4, we examined the luminosity--temperature ($L_{\rm X}$--$T_{\rm in}$) relation derived from the \textit{diskbb} fits (Fig.~\ref{Lx_Tin}). The data do not follow the canonical $L \propto T^{4}$ relation expected for a standard thin accretion disk with a constant inner radius \citep{2000ApJ...535..632M}. Instead, the measurements show significant scatter and do not reveal a statistically robust correlation when all epochs are considered. A Spearman rank correlation test between $T_{\rm in}$ and $L_{\rm X}$ yields $\rho=-0.80$ with $p=0.20$, suggesting a possible negative correlation between luminosity and temperature, although the trend is not statistically significant given the limited number of observations. Excluding observation C8, which has the largest uncertainties in $kT_{\rm in}$ due to low photon statistics, the remaining epochs suggest a trend that can be approximately described as $L_{\rm X} \propto T_{\rm in}^{-2.5}$. However, the currently available observations sample only a limited range of spectral states, and future observations may reveal a broader distribution of luminosities and disk temperatures. This behavior may suggest that the apparent inner disk radius is not constant and that the observed thermal emission does not originate from a standard thin accretion disk. In the framework of super-Eddington accretion, such deviations from the $L \propto T^{4}$ relation are commonly expected, as the emission may arise from an extended, optically thick photosphere associated with a radiatively driven wind. In this scenario, an increase in accretion rate leads to an expansion of the emitting region and a decrease in temperature. Given the limited number of observations, the statistical robustness of the correlation remains tentative. This behavior is broadly consistent with wind-dominated emission scenarios reported in several ULXs \citep{2007MNRAS.377.1187P,2009MNRAS.393L..41K,2015MNRAS.447.3243M}. In addition, no clear correlation between the power-law photon index and luminosity is evident from the current data. This may suggest that the spectral evolution of X--4 is not governed solely by changes in the accretion rate. Instead, additional effects such as variations in the accretion geometry, viewing angle, and radiatively driven outflows may also play an important role in shaping the observed X-ray spectrum.

The hardness--intensity behavior of X--4 reveals a complex spectral evolution across the \textit{Chandra}, \textit{Swift}/XRT, and \textit{XMM-Newton} observations (see Fig. \ref{F:hardness}). The \textit{Chandra} data show substantial hardness variations over a relatively broad flux range, with hardness ratios spanning $\sim0.3$--1.8, while the absorbed flux changes by nearly an order of magnitude. No clear monotonic correlation between hardness and flux is observed. Instead, observations with comparable fluxes can exhibit different hardness ratios. The \textit{Swift}/XRT observations probe the source during brighter states and are generally concentrated at comparatively high hardness ratios, indicating that the source predominantly remained in a relatively hard spectral state during the monitoring campaign. Although some hardness variability is present, the overall distribution does not reveal a clear monotonic relation between hardness and flux. In contrast, the \textit{XMM-Newton} observations are concentrated at comparatively low hardness ratios, indicating softer spectral states compared with most of the \textit{Chandra} and \textit{Swift}/XRT measurements. However, given the limited number of \textit{XMM-Newton} epochs, this result should be interpreted with caution and does not by itself establish the presence of a distinct long-term spectral state transition. Taken together, the hardness--intensity behavior of X--4 does not follow the canonical monotonic hard-to-soft evolution commonly observed in sub-Eddington Galactic XRBs (e.g., \citealp{2006ARA&A..44...49R}). In particular, some observations exhibit comparatively hard spectra at lower fluxes, whereas softer spectral states are also present at intermediate luminosities. This behavior suggests that the spectral evolution of X--4 is not governed solely by luminosity changes or variations in the mass accretion rate. Rather, the observed variability may also reflect changes in the accretion geometry, radiatively driven outflows, and viewing-angle effects, as commonly inferred in super-Eddington ULXs \citep{2013MNRAS.435.1758S,2021A&A...649A.104G,2023A&A...669A.130A}.

The absence of detected coherent signals in X--4 does not allow us to constrain the nature of the compact object. In particular, the lack of detected periodicity does not rule out a neutron star accretor, as such signals in ULXs can be transient, geometrically suppressed, or below the detection threshold depending on the accretion geometry and beaming effects \citep{2014Natur.514..202B,2016MNRAS.458L..10K,2019MNRAS.484..687M}. Several ULXs hosting neutron stars exhibit strong aperiodic variability and spectral properties similar to those observed here \citep{2019MNRAS.484..687M}, even when no coherent pulsations are detected. Overall, the spectral and timing properties of X--4, including its non-monotonic hardness evolution and short-timescale variability, are consistent with a super-Eddington accretion scenario onto a stellar-mass compact object, although the current data do not allow us to distinguish between a neutron star and a black hole accretor.

\subsection{Short and long-term variability}

X--4 exhibits substantial long-term X-ray luminosity variability across different observing campaigns. During the \textit{Swift}/XRT monitoring, the 0.3--10 keV luminosity varies by a factor of six. The variability amplitude increases significantly in the \textit{Chandra} observations, where the luminosity changes by a factor of 25 between the faintest and brightest detected states. In the \textit{XMM-Newton} observations, the luminosity varies by nearly two orders of magnitude. Considering all observations together, the source luminosity varies by more than two orders of magnitude, demonstrating large-amplitude long-term variability (see panel~a of Fig.~\ref{F:long_term}). This variability amplitude is consistent with that observed in many transient ULXs \citep{2019MNRAS.483.3566V,2020ApJ...891..153E,2021MNRAS.501.1002W,2023MNRAS.525.3330R,2023ApJ...951...51B,2023MNRAS.526.5765A}.

Peak-like structures were observed during epochs when the source luminosity remained well below the canonical ultraluminous regime ($L_{\rm X} < 10^{39}\ {\rm erg\ s^{-1}}$). These events appear as short-timescale count-rate peaks with characteristic durations of $\sim1$--$5$ ks. Despite the presence of these peaks, the spectral properties differ among the observations: C2 exhibits a comparatively softer spectrum ($H/S = 0.70 \pm 0.12$), whereas C6, C8, and C10 are characterized by harder spectral states with $H/S \sim 1$. No statistically significant spectral evolution is detected during the peaks themselves. In particular, the hardness ratios measured during the peaks are consistent with those of the surrounding bins within the large statistical uncertainties. The observed variability may be associated with changes in the optically thick wind or variations in line-of-sight obscuration, although the current data do not allow the energy dependence of the variability to be robustly constrained. Similar variability behavior has been reported in several ULXs and is commonly interpreted as arising from clumpy or inhomogeneous outflows crossing the line of sight (e.g., \citealp{2017MNRAS.468.2865P,2018MNRAS.479.3978K}).

Moreover, Table~\ref{T:var_test} summarizes the results of the $\chi^{2}$ and Gregory--Loredo (GLV) variability tests for the \textit{Chandra} and \textit{XMM-Newton} observations of X--4. The $\chi^2$ test indicates statistically significant variability in C2 ($p=10^{-4}$) and suggests likely variability in C8 ($p=0.03$), while the remaining \textit{Chandra} observations are formally consistent with constant emission at the adopted significance threshold ($p \geq 0.05$). In contrast, the GLV test identifies C2, C4, C6, C8, and C10 as definitely variable, and C7 as consistent with constant emission, highlighting the greater sensitivity of the unbinned method in low-count regimes. Notably, C10 is classified as constant by the $\chi^{2}$ test ($p=0.51$) but as definitely variable by GLV, likely reflecting the limited photon statistics and the reduced power of binned statistics in this observation. C9 remains unconstrained due to insufficient counts.

Both \textit{XMM-Newton} observations (XMM1 and XMM2) show highly significant variability ($p=10^{-9}$ and $10^{-4}$, respectively), indicating persistent aperiodic variability in the higher signal-to-noise data with no evidence for periodic, quasi-periodic, or flare-like events. The two \textit{Chandra} observations obtained on the same day (C10 and C11) may indicate rapid changes in the count-rate behavior in X--4 on sub-day timescales. C10 is detected at a lower count rate and shows indications of short-timescale variability, with a nominally harder spectrum ($H/S = 1.21 \pm 0.37$). In contrast, C11 is brighter and does not show significant intra-observation variability, with a nominally softer spectrum ($H/S = 0.75 \pm 0.10$). Although the hardness ratios are formally consistent within the uncertainties, the observed count-rate and variability differences may suggest changes in the accretion flow properties between the two observations. Within the framework of super-Eddington accretion, such behavior may be associated with variations in the geometry of the inner accretion flow and radiatively driven outflows \citep{2015MNRAS.447.3243M,2015MNRAS.454.3134M,2017MNRAS.468.2865P}.

\section{Conclusion and summary} \label{C}

We presented the first detailed spectral--timing study of the transient ULX NGC~4631~X--4 using archival \textit{Chandra}, \textit{XMM-Newton}, and \textit{Swift}/XRT observations. Our main results are summarized as follows:

\begin{enumerate}

\item X--4 exhibits strong long-term X-ray variability. The 0.3--10 keV luminosity varies by more than two orders of magnitude over the full observational baseline, confirming the transient nature of the source.

\item Several peak-like structures with characteristic durations of $\sim1$--$5$ ks are observed in the \textit{Chandra} light curves. However, owing to the limited photon statistics and the presence of peaks in both relatively harder and softer observations, the energy dependence of this variability cannot be robustly constrained.

\item No coherent pulsations, quasi-periodic oscillations, or statistically significant periodic signals are detected in the available \textit{Chandra} and \textit{XMM-Newton} data.

\item  The X-ray spectra can be adequately described by both absorbed \textit{diskbb} and power-law models, with characteristic disk temperatures of $kT_{\rm in}\sim0.85$--$1.37$ keV and relatively steep photon indices of $\Gamma\sim2$--2.4.

\item The luminosity--temperature relation and the non-monotonic hardness evolution do not follow the behavior expected for a standard thin accretion disk. The observed spectral evolution of X--4 can instead be interpreted within the framework of super-Eddington accretion flows, where radiatively driven winds and viewing-angle effects affect the observed spectral appearance.

\item The observed short-timescale variability, including the aperiodic variability detected in the \textit{XMM-Newton} observations together with the rapid peak-like structures observed in the \textit{Chandra} light curves, further supports a scenario in which the emission is shaped by a radiatively driven, optically thick wind.

\end{enumerate}

In summary, the spectral and timing properties of X--4 place it within the population of super-Eddington accreting ULXs, where the observed emission is governed by both the accretion rate and the viewing geometry. The inferred spectral parameters and characteristic disk temperatures favor a stellar-mass compact object, consistent with either a neutron star or a stellar-mass black hole accretor. Although no coherent pulsations are detected, the current data do not exclude a neutron star, as similar variability and spectral properties are observed in known neutron star ULXs. Future deeper observations with higher signal-to-noise ratio will be crucial to constrain the nature of the compact object and to probe the structure of the supercritical accretion flow.

\begin{acknowledgements}
We thank the anonymous referee for constructive comments and suggestions that helped to improve the quality of this paper. This work is partially supported by the Bundesministerium f\"ur Wirtschaft und Energie through the Deutsches Zentrum f\"ur Luft- und Raumfahrt e.V. (DLR) under the grant 50 OR 2517.
LD acknowledges funding from the Deutsche Forschungsgemeinschaft (DFG, German Research Foundation) - Projektnummer 549824807. VFS was supported by the German Research Foundation (DFG) grant WE\,1312/59-1.
\end{acknowledgements}

\bibpunct{(}{)}{;}{a}{}{,}
\bibliographystyle{ngc4631_x4}
\bibliography{ngc4631_x4}

\end{document}